\documentclass[12pt]{iopart}

\begin{document}

\title{Focus on Quantum Memories}

\author{Gavin Brennen}
\address{Centre for Engineered Quantum Systems, Department of Physics and Astronomy,
Macquarie University, North Ryde, NSW 2109, Australia}
\author{Elisabeth Giacobino}
\address{Laboratoire Kastler Brossel, Universit\'{e} Pierre et Marie Curie, Ecole Normale Sup\'{e}rieure and CNRS, UPMC, 4 place Jussieu, 75252 Paris Cedex 05, France}
\author{Christoph Simon}
\address{Institute for Quantum Science and Technology, Department of Physics and Astronomy, University of Calgary, Alberta, Canada T2N 1N4}
\begin{abstract}
Quantum memories are essential for quantum information processing and long-distance quantum communication. The field has recently seen a lot of progress, and the present focus issue offers a glimpse of these developments, showing both experimental and theoretical results from many of the leading groups around the world. On the experimental side, it shows work on cold gases, warm vapors, rare-earth ion doped crystals and single atoms. On the theoretical side there are in-depth studies of existing memory protocols, proposals for new protocols including approaches based on quantum error correction, and proposals for new applications of quantum storage. Looking forward, we anticipate many more exciting results in this area.
\end{abstract}

\maketitle

Just as classical computers are unthinkable without memories, quantum memories will be essential elements for future quantum information processors. Quantum memories for light are being studied particularly actively in the context of quantum communication for the implementation of quantum repeaters. Many different systems are being studied, including both cold and warm atomic gases, defect centers in solids, and individual atoms. The field of quantum memories has recently seen a lot of progress \cite{reviews}, with e.g. the achievements of very long storage times \cite{time}, high efficiencies \cite{eff}, various experiments involving entanglement \cite{entanglement}, and the realization of highly multimode memories \cite{multimode}.

The present focus issue offers a glimpse of this exciting field. It showcases both experimental and theoretical work from many of the leading groups in this area. On the experimental side it covers results in all of the systems mentioned above \cite{sparkes,clark,sprague,gundogan,sabooni,saglamyurek,huwer,kurz}. Ref. \cite{sparkes} reports on the implementation of a gradient echo memory in a cold atomic ensemble with very high optical depth. Also using the gradient echo memory protocol, Ref. \cite{clark} shows spatially addressable readout and erasure of an image in a warm atomic vapor. Ref. \cite{sprague} also uses a warm vapor, but inside a hollow-core photonic crystal fiber, demonstrating efficient optical pumping and high optical depth, which constitutes an important step towards implementing a Raman quantum memory in this system. There are three papers using the atomic frequency comb memory protocol in rare-earth ion doped crystals; Ref. \cite{gundogan} demonstrates coherent spin-wave storage of multiple temporal modes, Ref. \cite{sabooni} shows a large enhancement of the storage efficiency thanks to an optical cavity, and Ref. \cite{saglamyurek} uses the light-matter interface as an integrated processor for photonic quantum states. Finally there are two papers that represent important steps towards the heralded storage of photons in single atoms; Ref. \cite{huwer} demonstrates the detection of photon entanglement by a single atom, and Ref. \cite{kurz} uses an atom to realize a high-rate source of single photons, where the photons are suitable as heralds for single-photon absorption.

On the theoretical side, there are in-depth studies of existing quantum memory protocols \cite{hush,mendes,pascual}, proposals for new memory protocols \cite{vivoli,iakoupov,hetet,kaviani,afzelius,bombin,sarma,renes}, and proposals for new applications of quantum memories \cite{brunner,sangouard,clausen,moiseev}. As for new theoretical analysis of existing protocols, Ref. \cite{hush} uses a quantum input-output model to study gradient echo memories, which makes it possible to obtain analytical results in the regime of high-efficiency experiments. Ref. \cite{mendes} studies the reading process in quantum memories, providing an analytical description of the extracted photon and obtaining quantitative agreement with experiments. Ref. \cite{pascual} studies adiabatic rapid passage for rephasing in quantum memories, showing that it outperforms $\pi$ pulses.

As for new protocols, there are two papers making use of photon echo type phenomena; Ref. \cite{vivoli} proposes to use free induction decay in combination with spin wave storage to achieve very high time-bandwidth products, and Ref. \cite{iakoupov} proposes a modified version of the controlled reversible inhomogeneous broadening protocol that can achieve high efficiencies without any optical control fields. There are also two papers using polaritonic phenomena; Ref. \cite{hetet} proposes to control the splitting between two absorption lines to implement a new type of slow-light storage, whereas Ref. \cite{kaviani} proposes to sweep the frequency of the absorption line, leading to a protocol that has both slow-light and photon-echo type characteristics.
Ref. \cite{afzelius} proposes to realize a quantum memory for propagating microwave photons using a solid-state spin ensemble and a microwave cavity. Finally there are three proposals based on quantum error correction; Ref. \cite{bombin} proposes a model for self-correcting quantum memories and quantum computers (albeit in six spatial dimensions), Ref. \cite{sarma} studies a continuous-time version of stabilizer codes in nanophotonic circuits, and Ref. \cite{renes} studies holonomic quantum computing in ground states of spin chains.

In the category of proposals for new applications, there are two papers proposing to use heralded storage of photons in order to implement a loophole-free Bell test; Ref. \cite{brunner} proposes to use spin-photon interactions in cavities for this purpose, while Ref. \cite{sangouard} studies single atoms in free space. Ref. \cite{clausen} proposes to implement photon-number resolving photon detection based on light storage in trapped ion ensembles and fluorescence detection. Finally Ref. \cite{moiseev} studies photon waveform conversion in Raman echo quantum memories.

Going forward, we anticipate more exciting work on quantum memories with improved performance on key criteria such as efficiency, fidelity, storage time, and multimode capacity, and on achieving top performance for all of these benchmarks in a single system. Ever new challenges for experiments come from proposals such as quantum communication by ship \cite{ships} or quantum repeaters using satellite links \cite{sats}. We also expect more work on storing new degrees of freedom such as orbital angular momentum \cite{OAM}, on achieving ultrahigh bandwidth \cite{sussman}, and on storing microwave photons \cite{microwaves}. Finally we anticipate more work on using quantum memories for new purposes such as the implementation of photon-photon gates \cite{gates}, the creation of macroscopic superposition states \cite{lau}, and the development of entire quantum computing architectures \cite{architectures}. A major theoretical open question is the possibility of self-correcting quantum memories in a physically realistic architecture. It has been known for some time that a topological memory in $d=4$ spatial dimensions self-corrects up to a critical non-zero temperature \cite{Dennis}, and some models in three dimensions have relatively long lifetimes \cite{Haah}, however the possibility of scalable self correction at finite temperature in $d<4$ remains an outstanding problem.

\end{document}